\documentclass[journal,onecolumn]{IEEEtran}
\usepackage{amsmath,amsfonts}
\usepackage{algorithmic}
\usepackage{algorithm}
\usepackage{array}
\usepackage[caption=false,font=footnotesize,labelfont=sf,textfont=sf]{subfig}
\usepackage{textcomp}
\usepackage{stfloats}
\usepackage{url}
\usepackage{verbatim}
\usepackage{graphicx}
\usepackage{cite}
\usepackage{lipsum}
\usepackage{breqn} 
\usepackage{multirow}
\usepackage{tablefootnote}
\usepackage{hhline}
\usepackage[table,xcdraw]{xcolor} 
\usepackage{tablefootnote}
\usepackage{tabularx}
\usepackage{subfloat}

\begin{document}
\title{Calibration of Wideband LFM Radars based on Sliding Window Algorithm}
\author{Hyung-Woo Kim, Jin-woo Kim, Jin-ha Kim, JaeYoung Choi, Sangpyo Hong and Byungkwan Kim}
\maketitle
\thanks{ 
Corresponding Author: Byungkwan Kim (e-mail:byungkwan.kim@cnu.ac.kr) 
}%
\markboth{Journal of \LaTeX\ Class Files,~Vol.~14, No.~8, August~2021}%
{Kim \MakeLowercase{\textit{et al.}}: Calibration of Wideband LFM Radars based on Sliding Window Algorithm}

\vspace{0.5cm}
\begin{abstract}
	This paper addresses the challenges of wideband signal beamforming in radar systems and proposes a new calibration method. 
	Due to operating conditions, the frequency dependent characteristics of the system can be changed, and amplitude, phase, and time delay error can be generated.
	The proposed method is based on the concept of sliding window algorithm for linear frequency modulated (LFM) signals.
	To calibrate the frequency-dependent errors from transceiver and the time delay error from true time delay elements, the proposed method utilizes the characteristic of the LFM signal. 	
	The LFM signal changes its frequency linearly with time, and the frequency domain characteristics of the hardware are presented in time.
	Therefore, by applying matched filter to a part of the LFM signal, the frequency dependent characteristics can be monitored and calibrated.
	The proposed method is compared with the conventional matched filter based calibration results and verified by simulation results and beampatterns.
	Since the proposed method utilizes LFM signal as calibration tone, the proposed method can be applied to any beamforming systems, not limited to LFM radars.

\end{abstract}



\section{Introduction}

Radar performs various operations like surveillance, imaging, target detection tracking in remote location.
To perform these operations with high resolution and accuracy, radar system requires wideband signal and large number of transmit and receive modules (TRM) for beamforming\cite{skolnik1980introduction}.
The beamforming can be performed in time domain by true-time delay element (TTD) or in frequency domain by phase shifter. 
The phase shifter based beamforming, frequency-domain beamforming, is widely used in phased array radar system because of its lower cost than TTD elements\cite{sreenivasulu2021design}.
However, the frequency-domain beamforming has a limitation on the radar waveform, in terms of bandwidth and pulse duration, as follows.
\begin{equation}
	BW \times \tau_{sig} \leq 1
\end{equation}

If this limitation is not satisfied, the phase shifter based beamforming forms the beam in the undesired direction at some frequencies and the effective beam width is broadened.
Therefore, the time domain beamforming is required for the wideband beamforming radar systems.\cite{longbrake2011true} 
The required time delay for the beamforming is proportional to the distance between the phase center of antenna elements. 
For example, an S-band radar system with center frequency of 3.25GHz, the minimum delay between antenna is equivalent to 130ps.
Therefore, precise control of time delay between TRMs are required and true time delay elements are adopted for those requirements.

Recent phased array radar systems are designed to operate for multiple functions with various waveforms. 
Based on the target or operation mode, the radar system selects its waveform including bandwidth or additional binary phase modulation \cite{wang2018research}. 
This means that the radar system should be able to control its phase using the phase shifter, even it performs beamforming in time domain with TTD elements. 
On the other hand, the radar system can apply both time delay and phase shift for the beamforming by considering total cost and performance of the system \cite{guccione2022design}. 

The TRMs are composed of various components such as power amplifier, phase shifter, and many passive elements that have frequency-dependent characteristics\cite{maffei2020spaceborne}.
In narrowband assumption, the frequency-dependent characteristics are negligible and the TRMs are assumed to have constant characteristics in the frequency domain\cite{rau2000analytic}.
However, for wideband signal, the characteristics of the TRM are not constant in the frequency domain and the characteristics of the TRMs are different from each other.

Therefore, beamforming system requires precise control of the transmitted or received signal to have a desired differences in time, amplitude, phase in all frequencies.
To achieve this, many calibration methods have been proposed and implemented\cite{fujita1998polarimetric,buck2000asar,hounam2002active,boncori2008signal,schwerdt2009final,kim2015orbit}. 
The PN gating method is widely used technique for the calibration of the remote sensing radar system, which is based on the orthogonal code\cite{inproceedings,vollbracht2014system,brautigam2009efficient}. 
The calibration method can monitor characteristics of each TRMs by applying orthogonality of the waveform during the calibration process.
This method is simple and easy to implement, and can measure realistic characteristics of the TRMs since it monitors the characteristics of the TRMs when all TRMs are operating simultaneously.
However, the PN gating method is still targeted on the narrowband system and the calibration is performed with single value of amplitude, phase, and delay\cite{wang2009polarimetric}.

In this paper, we propose a monitoring and calibration method for wideband phased array radar system by application of sliding window algorithm to the chirp-pulsed signal.
The proposed method can monitor the characteristics of the TRMs in the frequency domain, by amplitude and phase of the TRMs, and in the time domain, by time delay of the TRMs.
This method does not alter any hardware of the radar system and can be applied to the existing radar system by proposed processing method.
The proposed method can be applied to the radar system with both TTD and phase shifter based beamforming system.

The rest of this paper is organized as follows.
In Section 2, the principles of the calibration of wideband phased array radar system with linear frequency modulation is described.
In Section 3, the proposed error monitoring and calibration method is presented.
In Section 4, the simulation setup, parameter, and results are presented.
Finally, the conclusion is given in Section 5.

\section{Calibration of Beamforming System with Linear Frequency Modulation}

The linear frequency modulation (LFM), also known as pulse compression technique, is a widely used waveform for the radar system to achieve high range resolution.
The LFM waveform is defined as follows. 
\begin{equation}
	x(t) = e^{j2\pi (f_{0}t + kt^2)}
\end{equation}
where $f_{0}$ is the center frequency of the LFM waveform and $k$ is the chirp rate of the LFM waveform.

The calibration of the beamforming system with LFM waveform is performed by the matched filter output of the received signal, 
to monitor the characteristics of the transmitter, receiver, cables, passive and active components.
The matched filter output represents relative delay, amplitude attenuation, and phase shift of the received signal compared to the reference signal\cite{wang2009polarimetric}.
This information is utilized to detect the target and to estimate the target location, or to generate the remote images of the target.
By using same principle, we utilize the matched filter output of the received signal to monitor the characteristics of the TRMs.

Since the beamforming is implemented by the relative difference in time and frequency domain of the signal,
the calibration is performed by setting a reference TRM and comparing the characteristics of the other TRMs to the reference TRM.
The signal delay can be measured by the index differences of the matched filter.
The amplitude attenuation and phase shift differences can be measured from the maximum value of the matched filter output, which means the calibration signal's amplitude and phase.
Further details of the calibration with LFM signal can be found in\cite{deng2011internal}. 

The conventional calibration of the beamforming system with LFM waveform is performed by the single value of amplitude, phase, and delay.
This means that if the TRM and any component has frequency dependent characteristics, the calibration result cannot be accurate.
To overcome this problem, we propose a new calibration method for wideband phased array radar system by adopting sliding window algorithm.

\section{Proposed Method}

\subsection{Frequency Dependent Error Model}
As described in the previous section, the conventional calibration can monitor and generate precise beampattern of the phased array radar system for narrowband signal.
However, if wideband signals are applied, the performance of the beamforming system can be different in frequency for various conditions.
The frequency-dependent performance variation of TRMs, phase shifter, TTDs can be found in literatures \cite{kumar2022demonstration}. 

Before describing the proposed method, we define our error model for the beamforming system.
Since the radar system considering in this study is based on the LFM waveform, the frequency error model can be defined also in time domain.
The frequency-dependent amplitude and phase can be represented in time valued function, since the LFM waveform changes its frequency linearly in time.
\begin{equation}
	E_{N}(t) = A_{N}(t)*exp(j2\pi*\phi_{N}(t))
\end{equation}\label{eq_lfm_signal}
where $A_{N}(t)$ and $\phi_{N}(t)$ are the amplitude and phase error of the $N$th TRM of beamforming system in frequency domain, respectively.
The time variable t can be converted to frequency variable $f$ by the linear relationship of LFM signal.
On the other hand, the relative delay error for any signal $s(t)$ can be represented as follows.
\begin{equation}
	s_{delayed}(t) = s(t-\tau_{N})
\end{equation}\label{eq_time_delay}
where $\tau_{N}$ is the relative delay error of the $N$th TRM of beamforming system.
Therefore, the frequency dependent error model of the beamforming system can be expressed as follows.
\begin{equation}
	x_{N}(t) = A_{N}(t-\tau_{N})*exp(j2\pi*\phi_{N}(t-\tau_{N}))*x(t-\tau_{N})
\end{equation}\label{eq_error_model}
where $x_{N}(t)$ is received signal including time delay, frequency dependent amplitude and phase errors.

In the simulation, the delay can be applied by shifting index of the signal, therefore, the precise delay error requires oversampling and decimation process of the signal.

\subsection{Calibration with Sliding Window Algorithm}

\begin{figure}[h]
	\centering
	\includegraphics[width=15 cm]{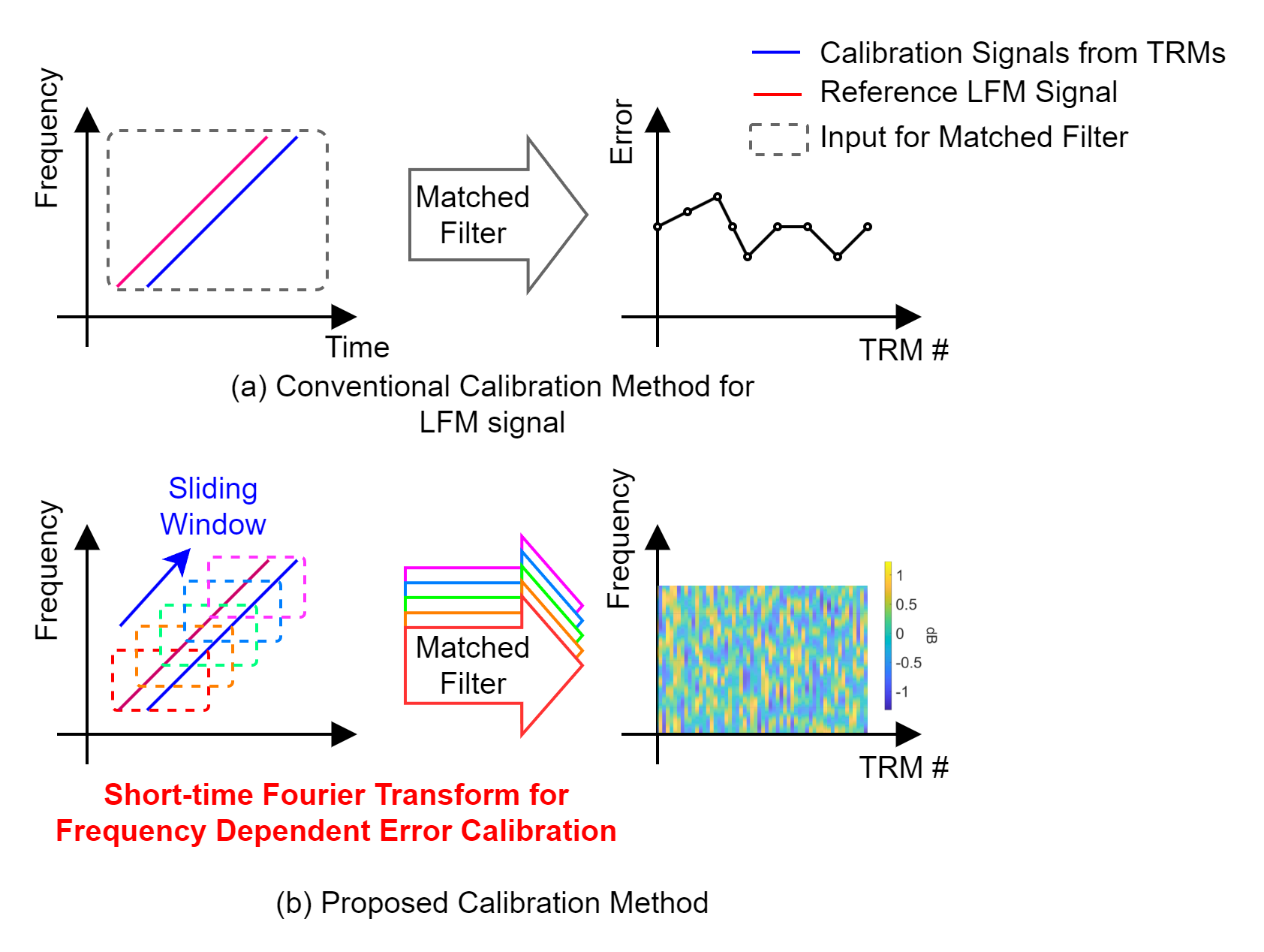}
	\caption{The conceptual diagram of the proposed method.\label{concept}}
\end{figure}

The proposed method is based on the sliding window concept for matched filtering, which is used to analyze the time-frequency characteristics of the signal.
The sliding window algorithm performs matched filter of the received LFM signal with a part of the calibration signal.
The main concept of the sliding window algorithm is the matched filtering of the LFM signals with different length, 
by dividing the LFM signal to represent a narrowband characteristic of the system.
The Fig. \ref{concept} shows the concept of the proposed method with sliding window algorithm.
The narrowband segmentation of the LFM signal is processed by the matched filter with same part of the reference signal.
By sliding the matched filter window in time domain, the characteristic of the LFM signal can be monitored in the frequency domain because the LFM signal is linearly changing its frequency.

\begin{figure}[!tbp]
	\centering
	\includegraphics[width=11 cm]{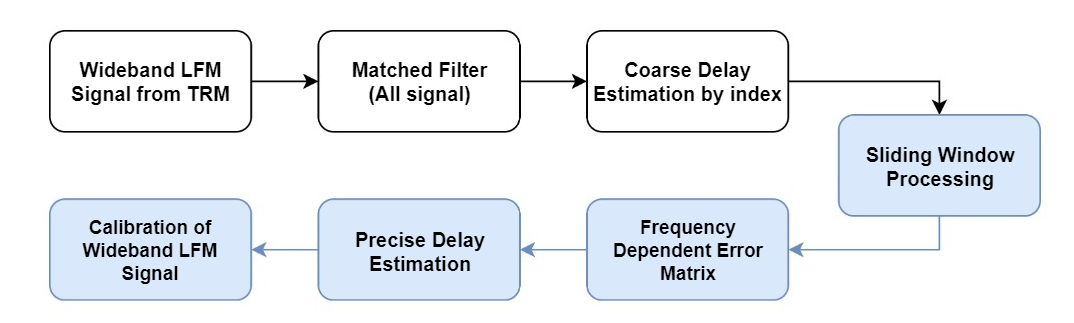}
	\caption{The procedure of proposed calibration method. The highlighted block is the proposed method, while normal blocks are conventional methods.\label{process}}
\end{figure}

The total process of the proposed method is shown in Fig. \ref{process}.
When the calibration process is started, the wideband LFM signal which includes the characteristic of each TRM is gathered.
The signal is processed with matched filter by the reference signal. 
A reference TRM is selected, and the coarse delay is estimated by the index difference of the maximum of matched filter output.
The coarse delay is compensated by shifting the index of the total signal.
After coarse delay compensation, the proposed method is started. The sliding window algorithm is performed to monitor the frequency dependent characteristics of the TRMs.
The amplitude and phase error for a certain frequency can be measured by the maximum value of the matched filter output from the LFM segmentation by a window.
Based on the measured error values, the proposed method constructs a error matrix in frequency domain.
The calibration of the TRMs can be performed by applying the error matrix to the received signal.

The calibration of the wideband LFM signal is performed by the same principle of the error model described in the previous section.
First, the time delay calibration is implemented by the shifting the total signal index by the measured delay error.
However, this calibration is not enough to compensate the precise time delay of the TRMs, since the time delay error is represented in the units of sample index of the signal.
For example, by supposing 1GHz bandwidth of the LFM signal with 10GHz center frequency, the sampling rate of the signal processing unit should be larger than 2GHz\cite{guccione2022design}. 
Assuming that the sampling rate of analog-digital converter is 2GHz, then the sampling time of a sample is 0.5ns, which means that the time delay error can be represented in the units of 0.5ns.
However, the beamforming radar will control the time delay in units of few picoseconds to generate high resolution beampattern, which means that the time delay error in the units of picoseconds should also be calibrated.
Therefore, the index based delay calibration is named as coarse delay calibration, and precise time delay calibration is separately performed in this study. 
The coarse delay calibration can be enhanced by increasing the sampling rate of the analog-digital converter, 
but it is not practical in the real system since the beamforming system should process large amount of digital signal in real time.

\begin{figure}[!tbp]
	\centering
	\includegraphics[width=8 cm]{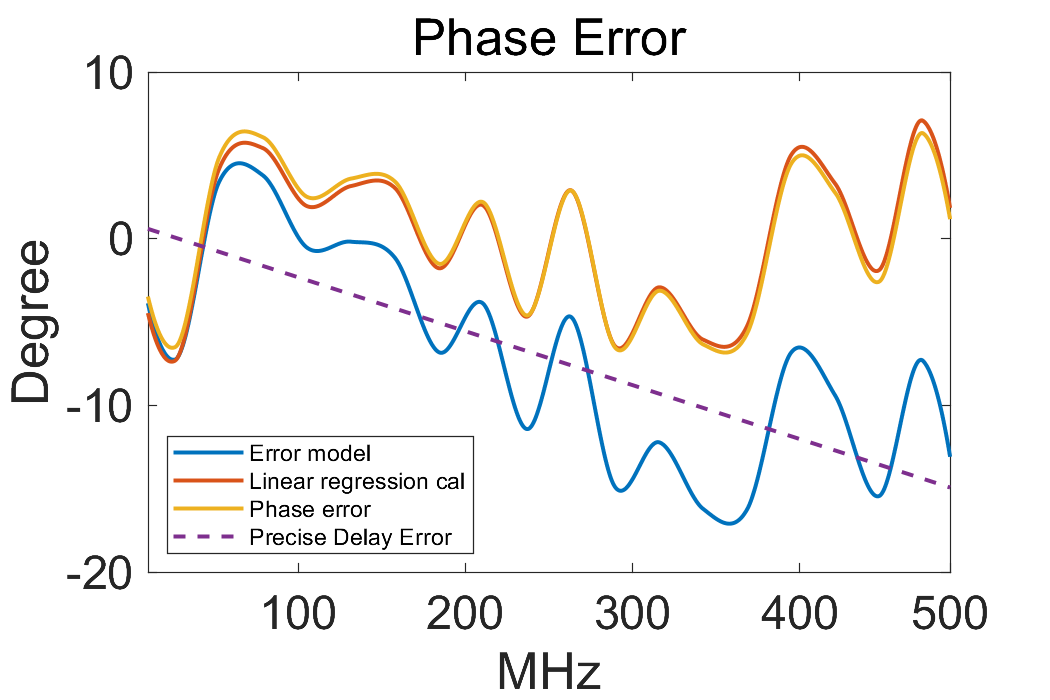}
	\caption{An example of frequency dependent phase error. Due to the physical nature of the microwave circuits, frequency dependent phase errors are generated. The linear phase error is from the time delay error. \label{precise_delay}}
\end{figure}

After coarse delay calibration, the signal still contains frequency dependent amplitude and phase errors, and precise time delay error. 
The frequency dependent amplitude and phase errors can be calibrated with the same method of the error model described in the previous section.
The phase error in this stage is consist of two parts, which are true frequency dependent phase error of the TRM and the linear phase term due to precise time delay.
The Fig. \ref{precise_delay} shows an example for the phase error consist of two parts.

The calibration of precise time delay can be done with two different approaches.
First method is the calibration of true-time delay element with the precise delay error by applying linear regression of measured phase error. 
This is effective only when the frequency-dependent phase error is very smaller than the additional linear phase due to time delay error. ($\phi_{N} << \tau_{N}*f(t)$)

If the beamforming radar system works fully in digital domain, the precise time delay can be considered as another phase error source that needs calibration.
In this situation, the second method can be introduced, the predistortion of the LFM signal with phase error, for the precise time delay and phase error of TRMs.
This can be implemented without additional computational cost, since the calculation is equivalent with proposed amplitude and phase calibration.

The proposed method can calibrate this phase error and precise time delay simultaneously since both errors are represented in the measured phase error.
Therefore, frequency dependent amplitude, phase error, and precise time delay of the $N$th TRM can be calibrated by the following equation. 

\begin{equation}
	x_{calibrated}(t) = x_{N}(t)*\frac{1}{A_{N}(t)}*exp(-j2\pi*\phi_{N}(t))
\end{equation}\label{eq_calibrated}

\section{Simulation \& Results}

\subsection{Simulation Parameters}

To verify the proposed method, the simulation is performed by considering realistic simulation parameters and beamforming system.
The parameters for error models are summarized in the following Table 1.
The wideband beamforming radar system which is based on both phase shifter and true time delay is considered in the simulation, as presented in \cite{guccione2022design}. 
The beamforming system consists of 64 TRMs, and the LFM signal is generated with 0.5GHz bandwidth, with 8us pulse length and 3.25GHz center frequency. 
\begin{table}[H] 
	\caption{The waveform and beamforming system parameters.}
	\label{table_system}
	\newcolumntype{C}{>{\centering\arraybackslash}X}
	\centering
	\begin{tabular}{c|c|c}
	\textbf{Parameters}	& \textbf{Value}	& \textbf{Unit}\\
	\hline
	Central Frequency & 3.25  & GHz  \\
	\hline
	Bandwidth         & 0.5   & GHz  \\
	\hline
	Pulsewidth        & 8     & us   \\
	\hline
	No. of TRMs       & 64    &      \\
	\hline
	No. of TTDs       & 8     &      \\
	\hline
	\end{tabular}
\end{table}

\subsection{Error model}
The error model of the beamforming system is generated by referencing measured error values of amplitude, phase, and time delay error from literatures.
The frequency dependent amplitude and phase error is modeled with interpolation from random values from the reported typical error range\cite{jeong20186}. 
The example of amplitude and phase error versus frequency for a TRM is shown in Fig. 4.
Each frequency dependent error model are generated and assigned to each TRMs by a frequency-amplitude matrix and frequency-phase matrix, as presented in the Fig. 11.
\begin{table}[H] 
	\caption{The amplitude, phase, and time error values for the simulation. The error values are selected to simulate the error levels similar with single bit range of phase shifter, true-time delay element. \label{table_error}}
	\newcolumntype{C}{>{\centering\arraybackslash}X}
	\centering
	\begin{tabular}{c|c|c}
	\textbf{Parameters}	& \textbf{Value}	& \textbf{Unit}\\
	\hline
	Amplitude Error              & -1.1 $\sim$ 0 & dB     \\
	\hline
	Phase Error                  & -7 $\sim$ 7     & degree \\
	\hline
	Coarse Time Delay Error Unit & 100              & ps     \\
	\hline
	Precise Time Delay Error Unit     & 20   & ps     \\
	\hline
	\end{tabular}
\end{table}
The time delay error is also referred typical error range of true time delay elements for the beamforming system\cite{jeong20186}. 
In digital domain, the time delay can be assigned to a signal by shifting the index of the signal. Therefore, the oversampling and decimation process is introduced to represent the time delay error lower than the sample time.
In this simulation, we assigned time delay with 20ps step by 10GHz oversampling, and all simulations are performed after decimating the signal with 5.
Therefore, the coarse delay error can be represented with 0.1ns step, and the precise time delay error is represented with 20ps step in the simulation.

By applying these error model to reference LFM signal, the error signal with frequency-dependent and time delay can be generated.

\subsection{Scenario and Results} 

\subsubsection{Error Measurement}

In this simulation, we assign errors to selected TRMs to verify the error measurement performance of proposed method.
To measure the various types of error, we assigned following different error model to the TRMs.
Conventional and Proposed calibration method are performed to the error signal, and the error values are compared with the assigned error values. 

Three different types of error are prepared for the simulation to verify the proposed method and compare with the conventional method.
The first type of error is the frequency constant amplitude and phase error, which is the most common error model for the narrowband beamforming system.
The second type of error is the frequency dependent amplitude and phase errors from the electromagnetic components, 
and the third type of error is the frequency dependent amplitude and phase error with time delay error from the true time delay element.

For clear comparison, the error values are assigned to the TRMs as follows.
The type 1 error model is assigned to the TRM 10, and the type 2 error model is assigned to the TRM 20, and the type 3 error model is assigned to the TRM 30.

\begin{itemize}
	\item Frequency constant amplitude and phase error
\end{itemize}

\begin{figure}[H]
	\centering
	\includegraphics[width=11 cm]{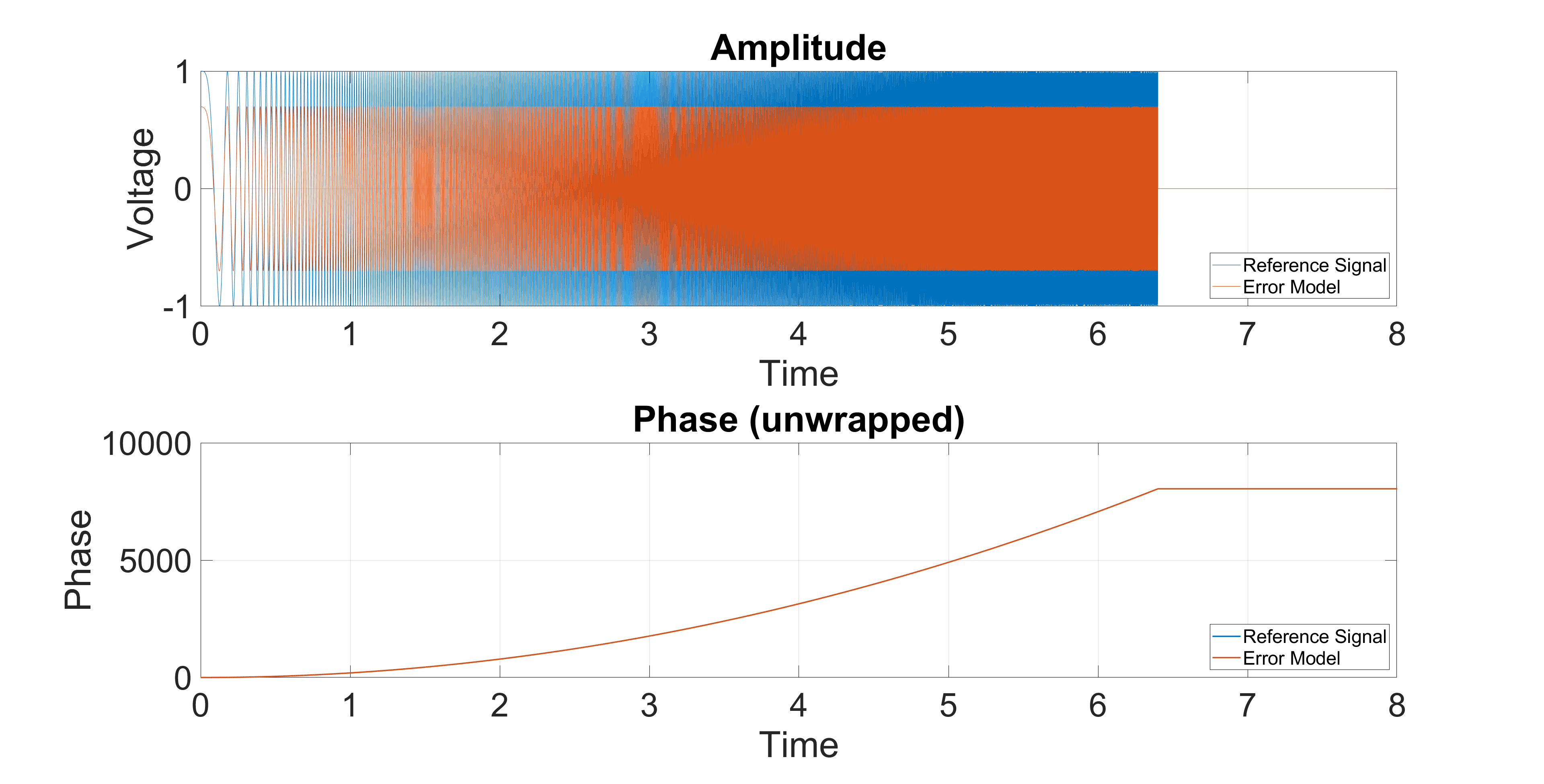}
	\caption{The amplitude and phase of frequency constant error applied signal.}
	\label{error_model_first_signal}
\end{figure}   

The amplitude error is assigned as -3dB, and the phase error is assigned as 5 degree.
The error signal is generated by multiplying the error model to the reference signal, and the amplitude and phase of the error signal is shown in Fig. \ref{error_model_first_signal}.

\begin{figure}[H]
	\centering
	\includegraphics[width=11 cm]{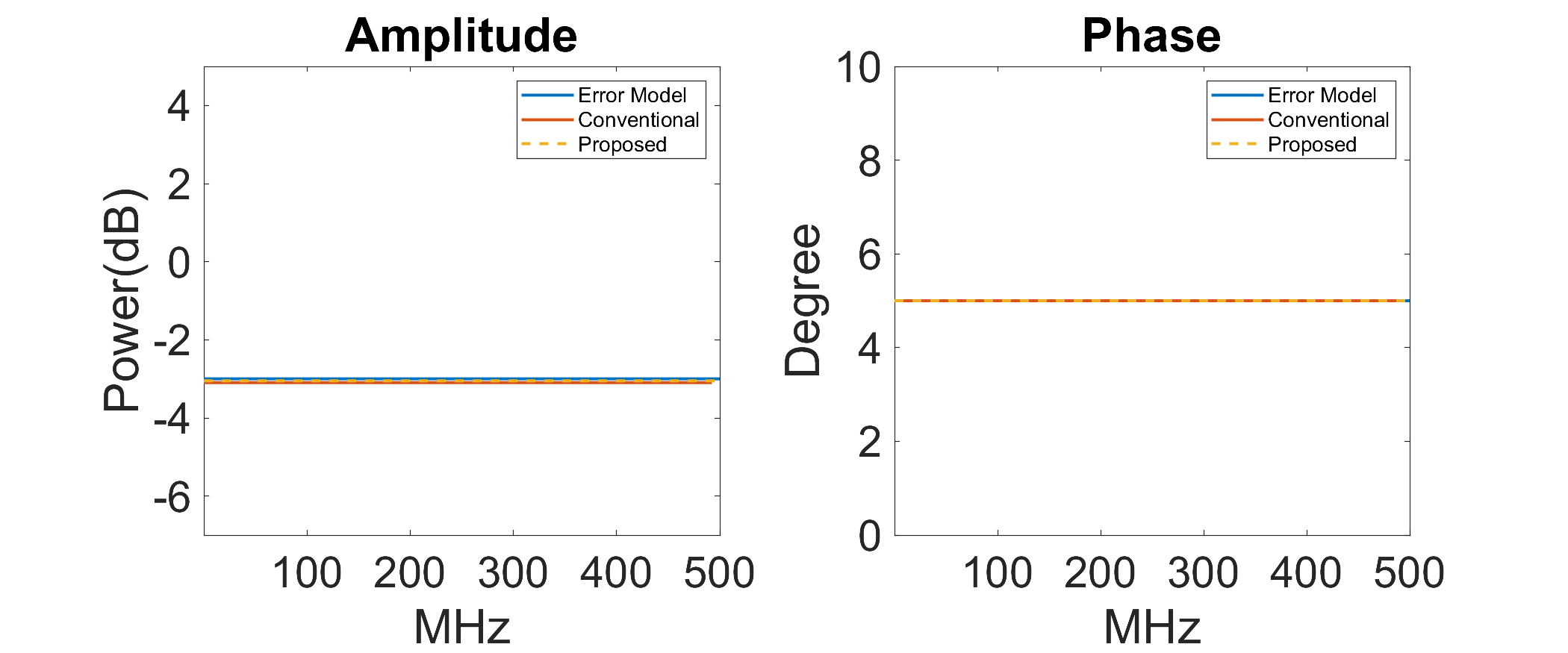}
	\caption{The assigned amplitude and phase error are presented with the results from the conventional method and the proposed method.}
	\label{error_model_first}
\end{figure}   

After calibration with the conventional and proposed method, the result error values are compared with the assigned error values for the frequency of LFM signal, as shown in Fig. \ref{error_model_first}.
Since the conventional method presents the error values with the same value for all frequency, the error values are presented as a line.
Both method presents the error values with the same value for the amplitude and phase, which means that the conventional method can calibrate the LFM signal when the error is constant for all frequency.
In this case, the proposed method can also calibrate the error, but the computational complexity is higher than the conventional method since it generates same values for all frequency.

\begin{itemize}
	\item Frequency dependent amplitude and phase error
\end{itemize}

\begin{figure}[H]
	\centering
	\includegraphics[width=11 cm]{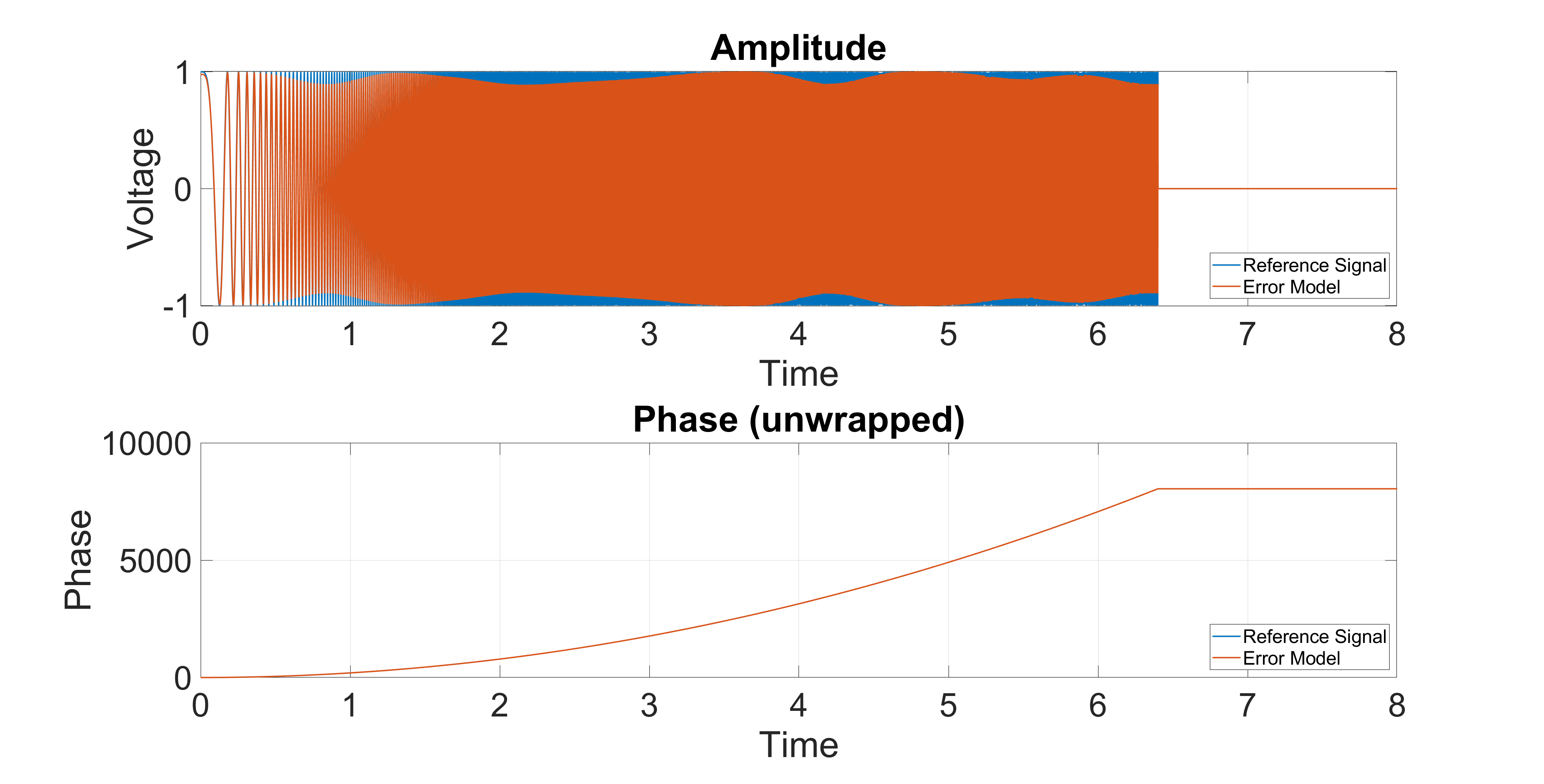}
	\caption{The amplitude and phase of frequency dependent amplitude and phase error applied signal.}
	\label{error_model_second_signal}
\end{figure}   

The frequency depedent error is assigned as the function of frequency, which is generated by interpolating the random values from the reported typical error range.
An example of the error applied signal is shown in Fig. \ref{error_model_second_signal}.

\begin{figure}[H]
	\centering
	\includegraphics[width=11 cm]{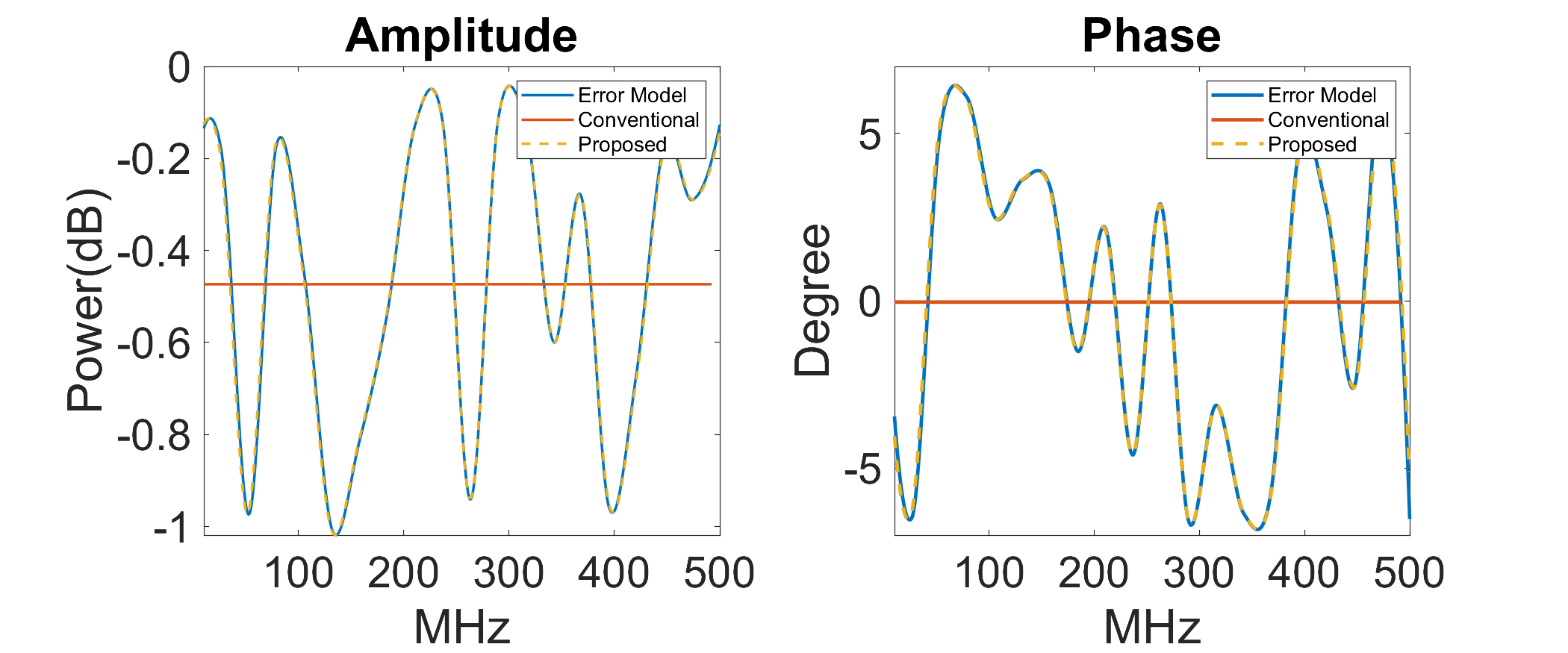}
	\caption{The assigned amplitude and phase error are presented with conventional and proposed calibration results.}
	\label{error_model_second}
\end{figure}   

The calibration result of error signal presented in Fig. \ref{error_model_second_signal} is shown in Fig. \ref{error_model_second}.
In the same manner, the error values from the conventional calibration are presented as a line because it does not consider frequency dependent error.
On the other hand, the proposed method presents the error values for each frequency bin. The calculated calibration accuracy of amplitude error of 0.016dB, and the phase error of 3.523degree.
The proposed method can calibrate the frequency dependent error, but the selection of window size and sliding step size is an important factor to the calibration accuracy.

\begin{figure}[H]
	\centering
	\includegraphics[width=11 cm]{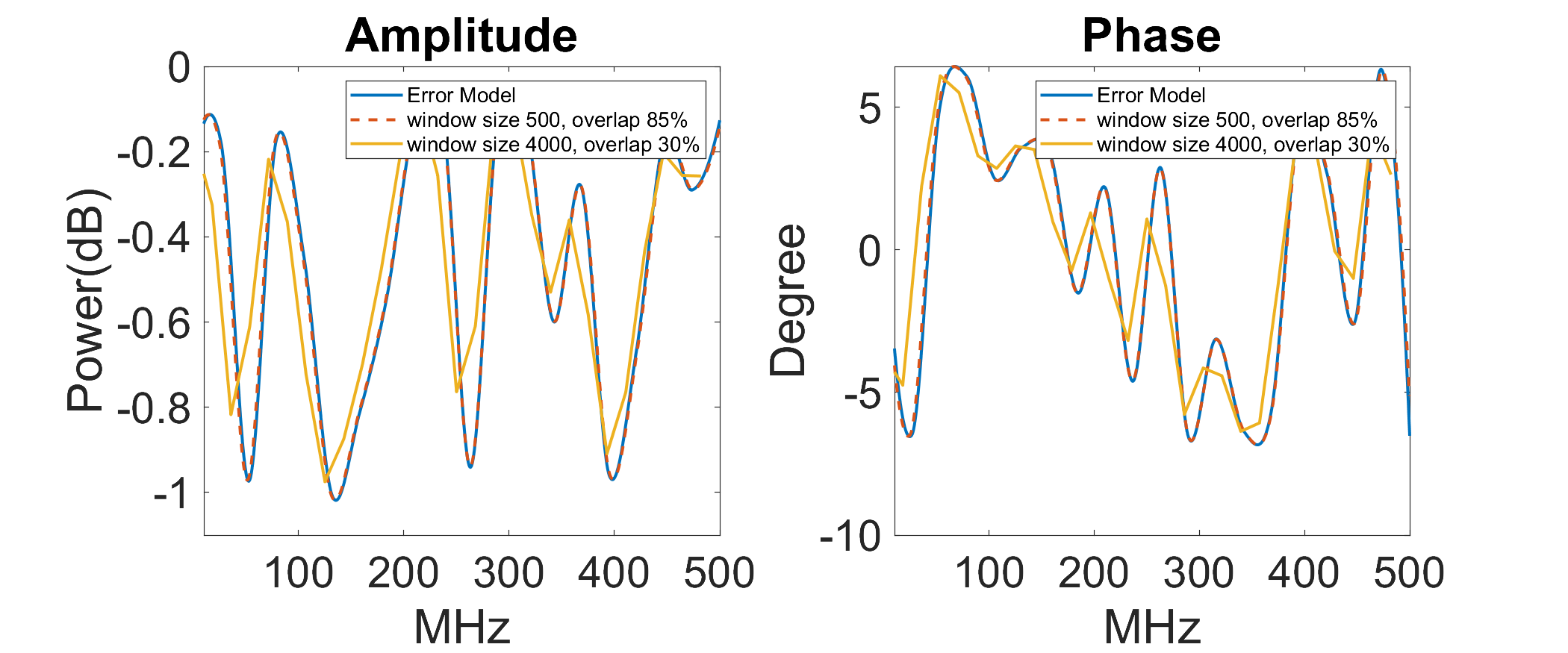}
	\caption{The proposed calibration result with different window size and overlapping ratio. The smaller window size and larger overlapping ratio shows higher calibration accuracy.}
	\label{error_model_second_window}
\end{figure}   

To show the effect of window size and sliding step size, the calibration results are shown in Fig. \ref{error_model_second_window}.
The window size is set to 500 and 4000, which slices 50ns and 400ns length of the LFM signal.  
The sliding step is determined by the overlapping ratio, which is set to 85\% and 30\%, 75 and 2800 samples for 500ns and 4000ns window size, respectively.
Both parameters determines the number of error values versus frequency, which is directly related to the calibration accuracy.
This is due to the nature of the LFM signal, which has the linear frequency increasement in time.
The smaller window size will represent the narrowband characteristics, 
and the larger overlapping ratio will contribute the changing characteristics of the error values in frequency domain.
Therefore, the larger window size and smaller overlapping ratio shows lower calibration accuracy as shown in Fig. \ref{error_model_second_window}.

\begin{itemize}
	\item Frequency dependent amplitude and phase error with time delay error
\end{itemize}

\begin{figure}[H]
	\centering
	\includegraphics[width=11 cm]{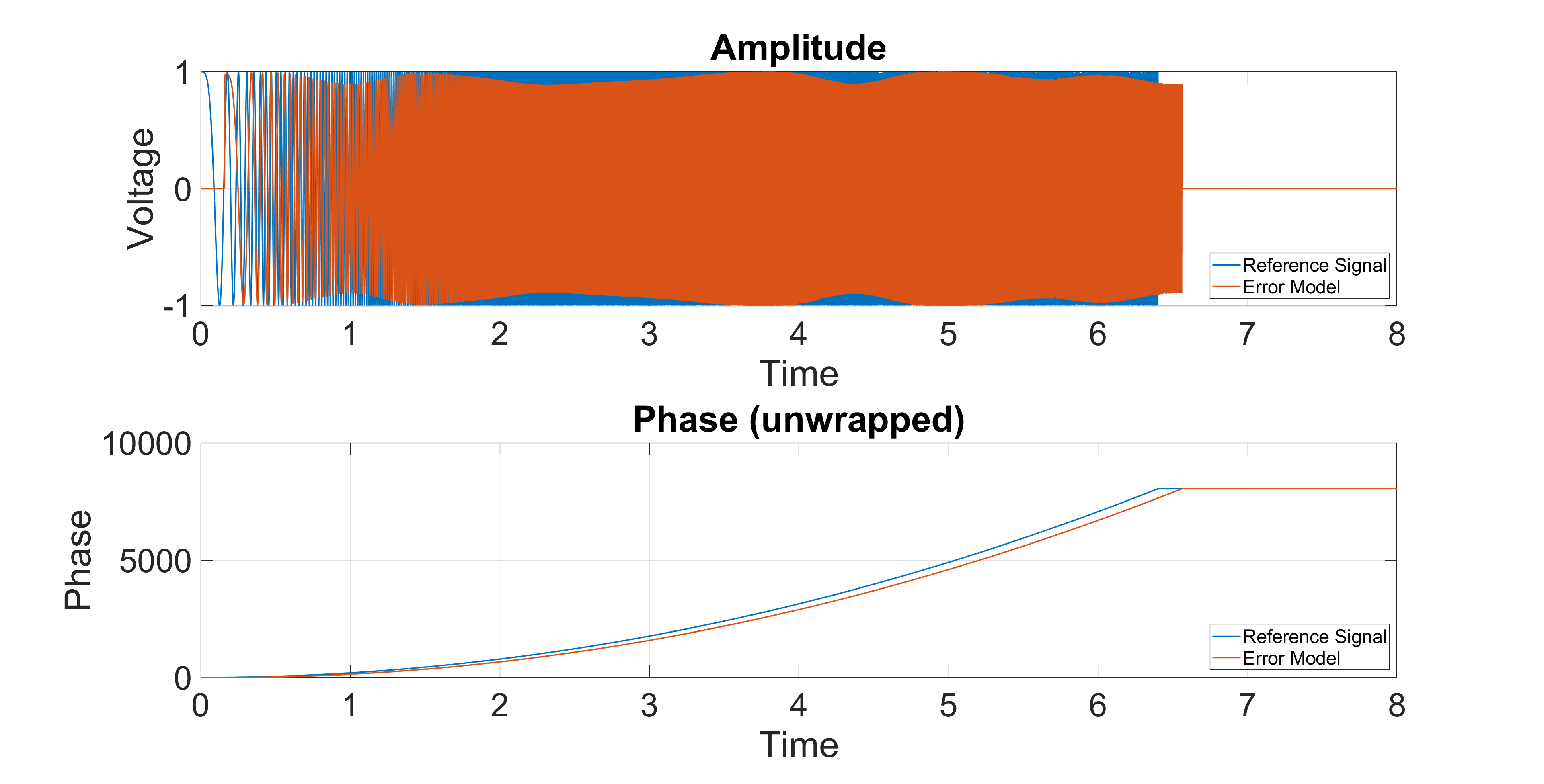}
	\caption{The amplitude and phase of frequency dependent amplitude, phase error and time delay error applied signal example. We applied 100ns delay for better presentation of time delayed signal. Due to the time delay error, the phase difference of the error signal is increasing with time. }
	\label{error_model_third_signal}
\end{figure}

As decribed in the proposed calibration process, the coarse delay error is calibrated first, and the precise delay error is calibrated after the coarse delay error is calibrated.
Therefore, the phase error from the precise delay error is included in the error signal as presented in Fig. \ref{precise_delay}, and it needs to be calibrated with the proposed method.
In this simulation, the total delay error is set to 100 ps and calibrated coarsely, and the precise delay error is set to 20 ps.
The error signal with the precise delay error is shown in Fig. \ref{error_model_third_signal}.
The Fig. \ref{error_model_third_signal} shows that the phase difference of the error signal is increasing with time, which is from the precise delay error.

\begin{figure}[H]
	\centering
	\includegraphics[width=11 cm]{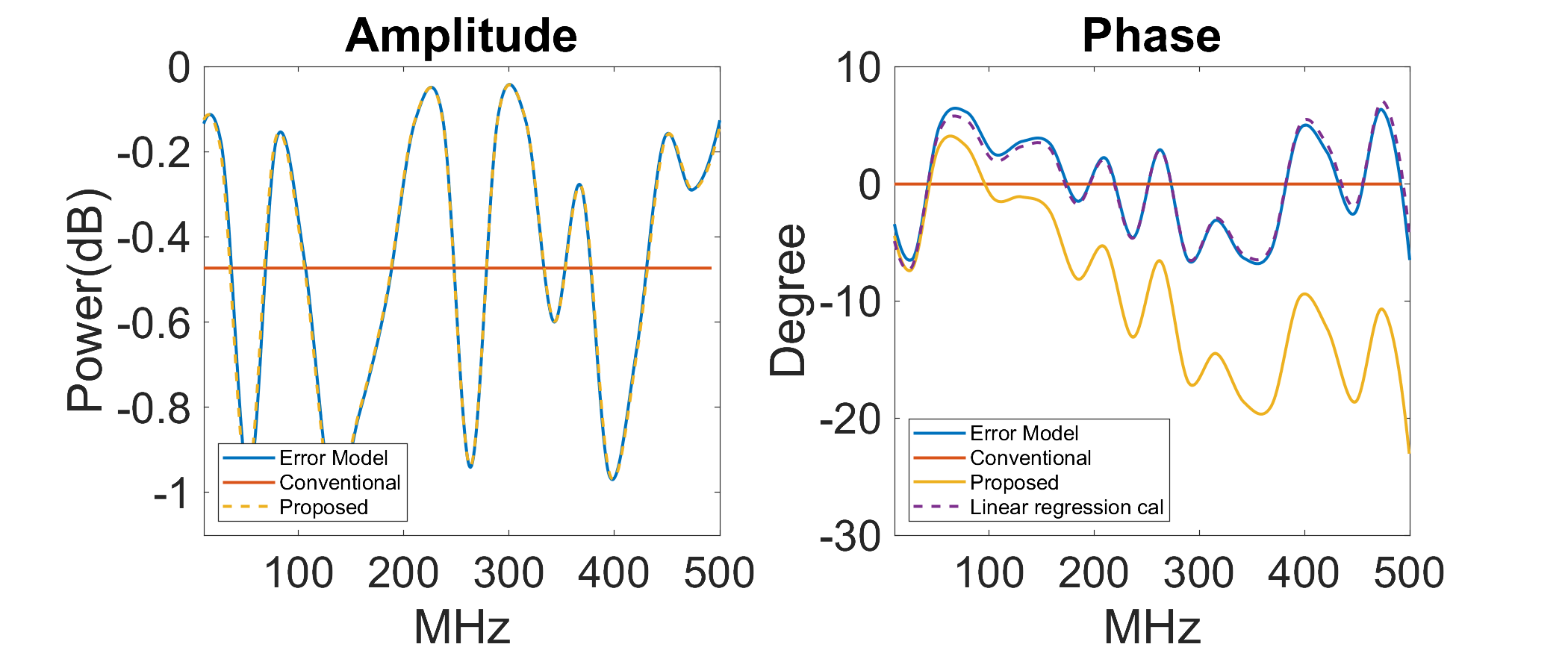}
	\caption{The assigned amplitude and phase error are presented with conventional and proposed calibration results.}
	\label{error_model_third}
\end{figure}

The calibration result of error signal with precise time delay is presented in Fig. \ref{error_model_third}.

From the calibration result, the precise time delay does not affect the calibration accuracy of amplitude error, but it affects the calibration accuracy of phase error.
As presented in Fig. \ref{error_model_third}, the applied phase error is added with a linear function of frequency.
By applying simple linear regression model to the phase error, the slope of the phase error is calculated as -3 degree/GHz, equivalent to 16 ps of precise delay.
The difference of the precise delay error applied and obtained is 2 ps, which is below the requirement of true time delay element resolution.

\begin{figure}[H]
	\centering
	\subfloat[][Applied Frequency Dependent Amplitude Error Model.]{\includegraphics[width=5.5 cm]{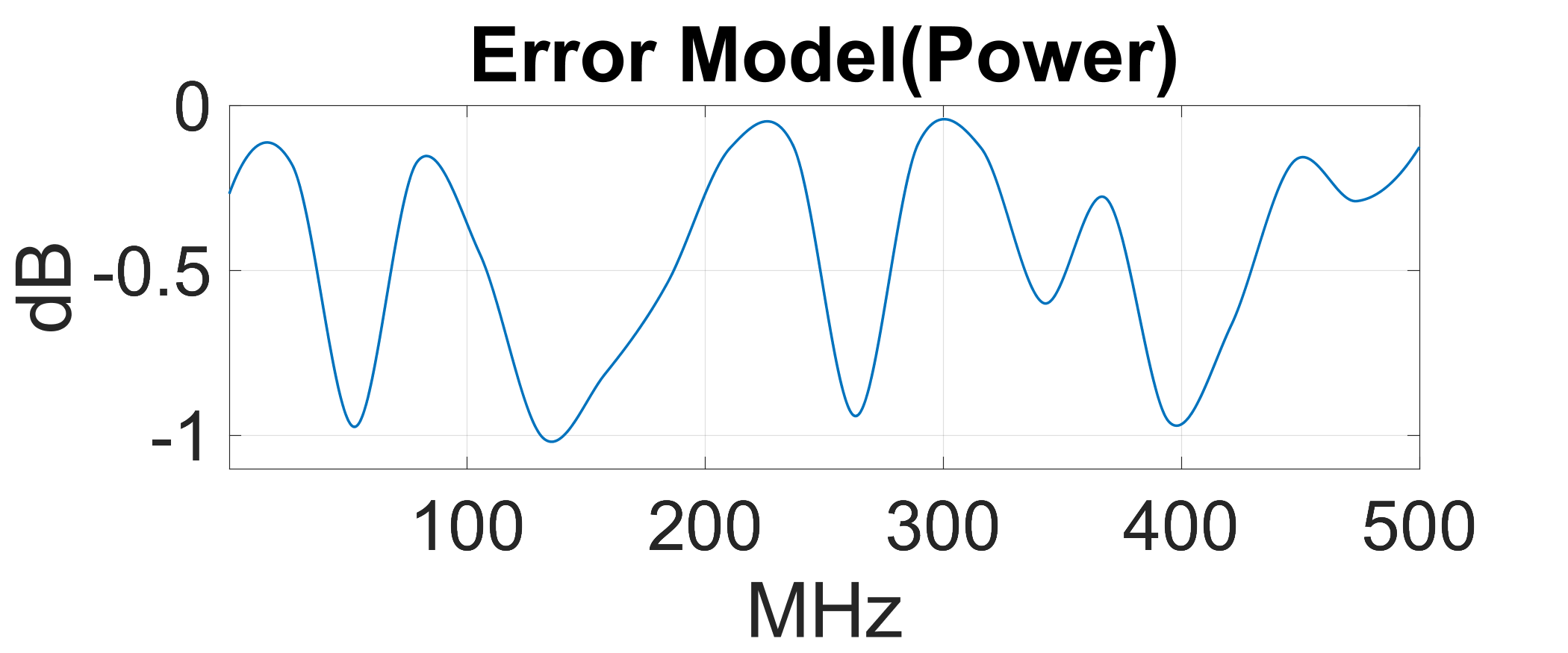}\label{freq_vs_amplitude}}
	\hspace{1em}
	\subfloat[][Applied Frequency Dependent Ph-ase Error Model.]{\includegraphics[width=5.5 cm]{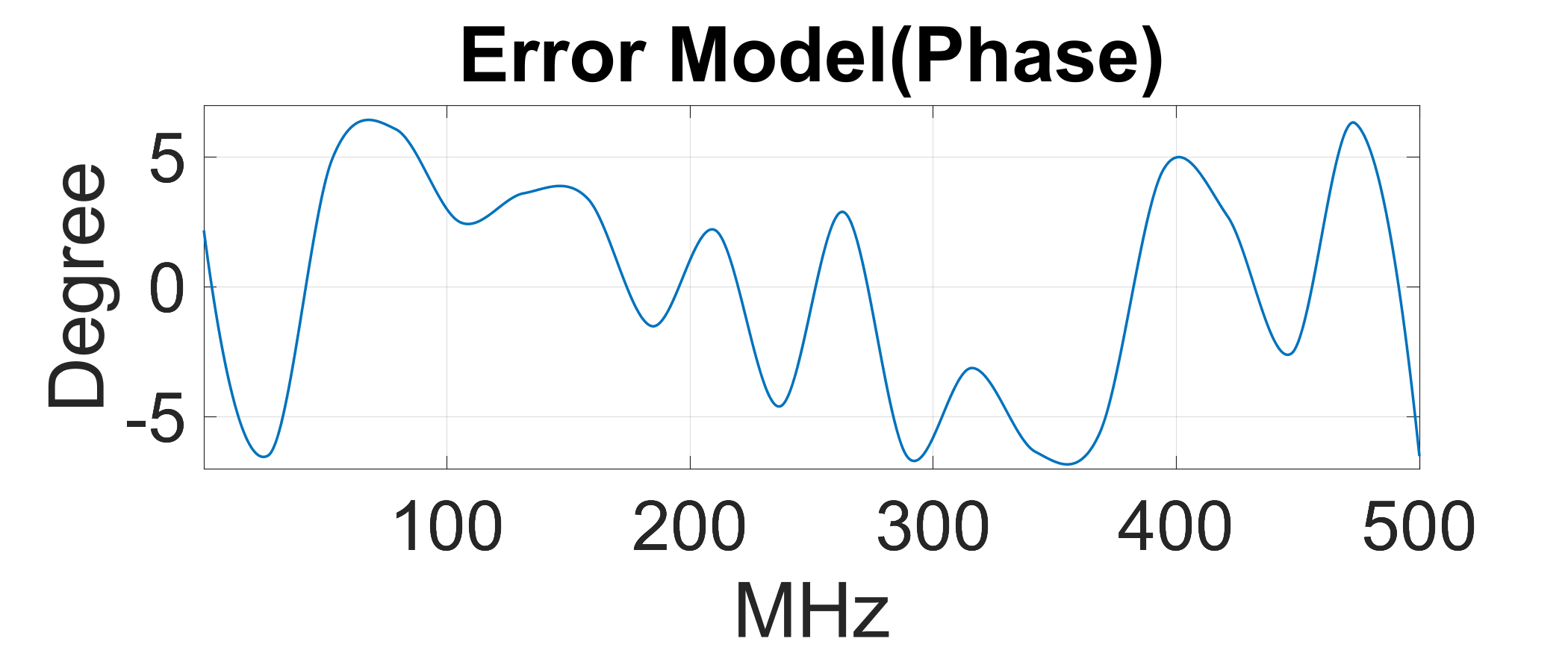}\label{freq_vs_phase}}
	\vspace{0.1em}
	\centering
	\subfloat[][Amplitude Calibration Values with Conventional Method.]{\includegraphics[width=5.5 cm]{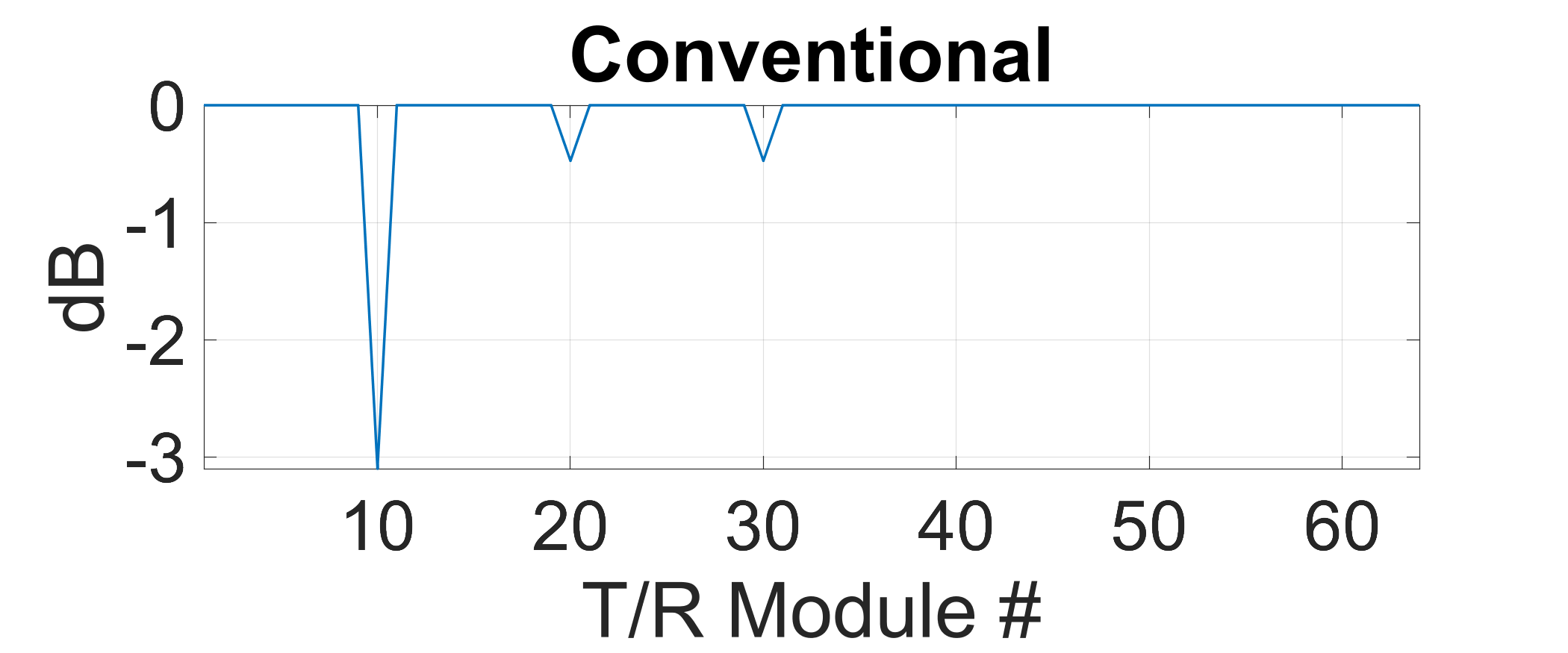}\label{conventional_amplitude}}
	\hspace{1em}
	\subfloat[][Phase Calibration Values with Conventional Method.]{\includegraphics[width=5.5 cm]{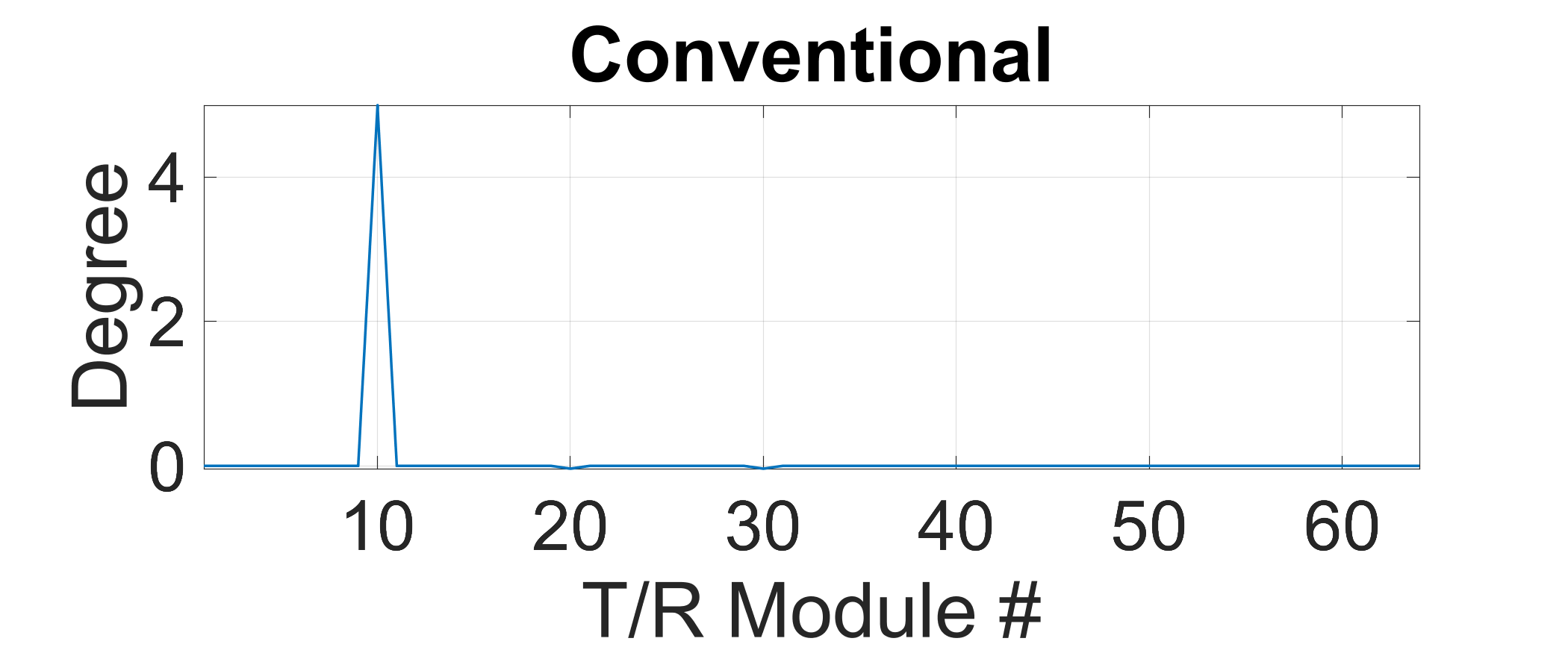}\label{conventional_phase}}
	\vspace{0.1em}
	\centering
	\subfloat[][Amplitude vs. Frequency Calibration Matrix with Proposed Method.]{\includegraphics[width=5.5 cm]{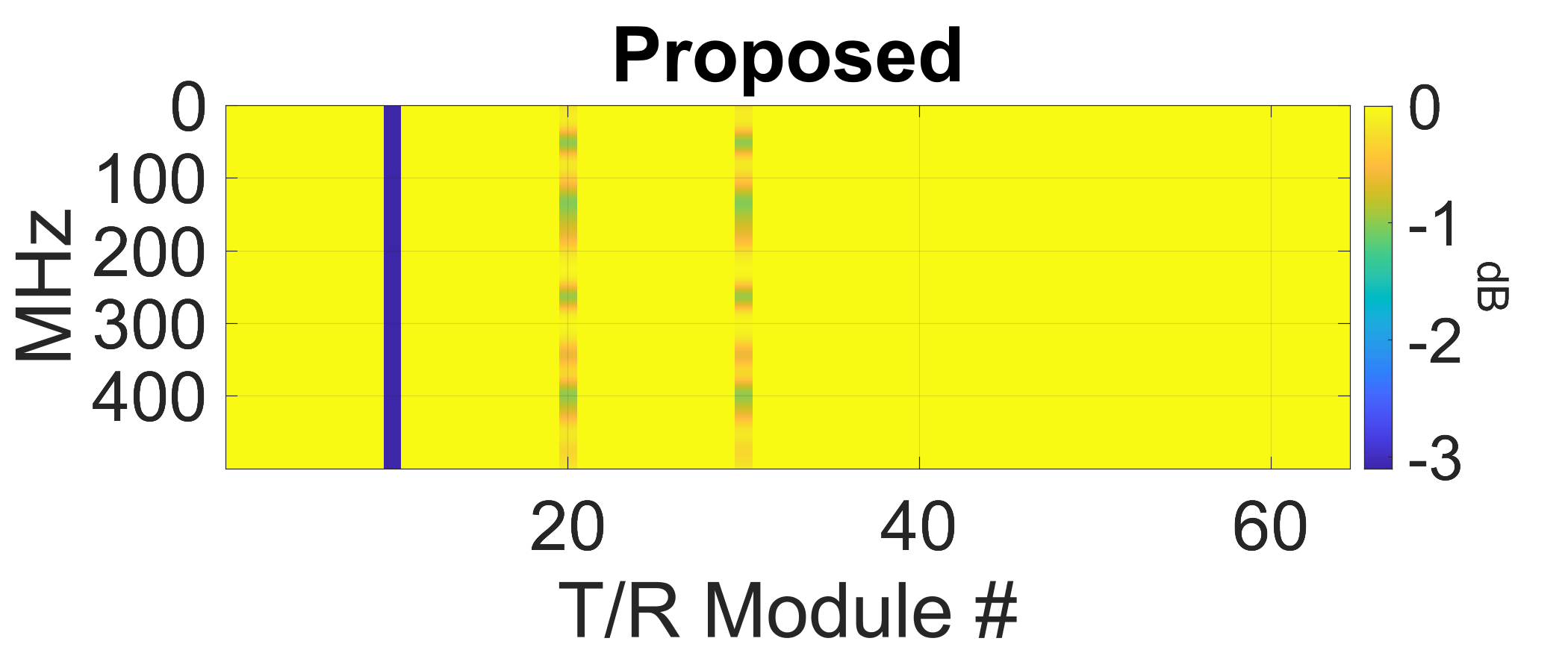}\label{proposed_amplitude}}
	\hspace{1em}
	\subfloat[][Phase vs. Frequency Calibration Matrix with Proposed Method.]{\includegraphics[width=5.5 cm]{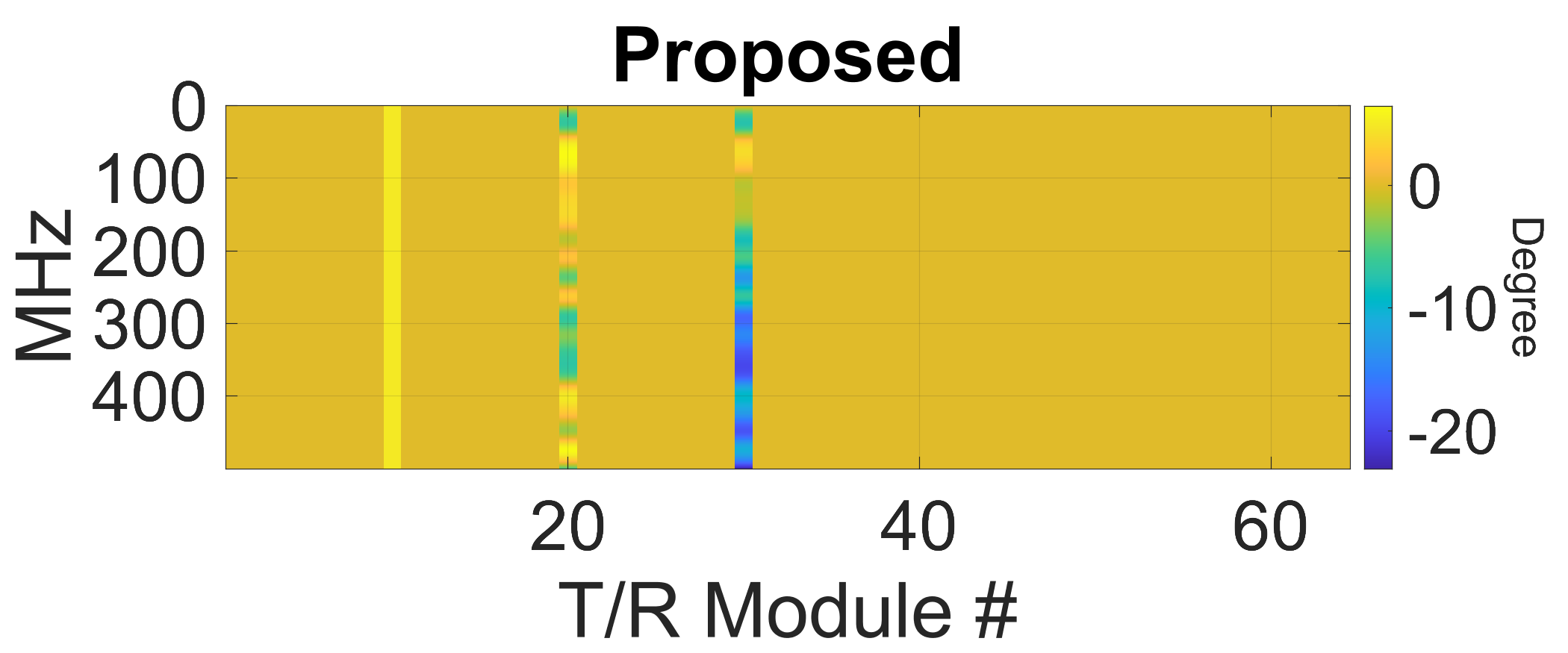}\label{proposed_phase}}
	\caption{The TRM number 10, 20, and 30 are assigned with the constant amplitude and phase error, frequency dependent amplitude and phase error, and frequency dependent amplitude and phase error with precise time delay error, respectively. The conventional and proposed calibration method are performed to the find out error from reference LFM signal.}
	\label{calibration_values_102030}
\end{figure}

In summary, the proposed calibration method is compared with the applied error model and the conventional calibration method for three different error types, presented in Fig. \ref{calibration_values_102030}.

Therefore, the proposed calibration method can calibrate the frequency dependent amplitude and phase error with precise time delay error.
In next section, the calibration result of the conventional and proposed method is utilized to generate the beamforming result.

\subsubsection{Calibration Effects on Beamforming}

\begin{figure}[H]
	\centering
	\subfloat[][The amplitude errors for each TRM by the conventional calibration method.]{\includegraphics[width=6 cm]{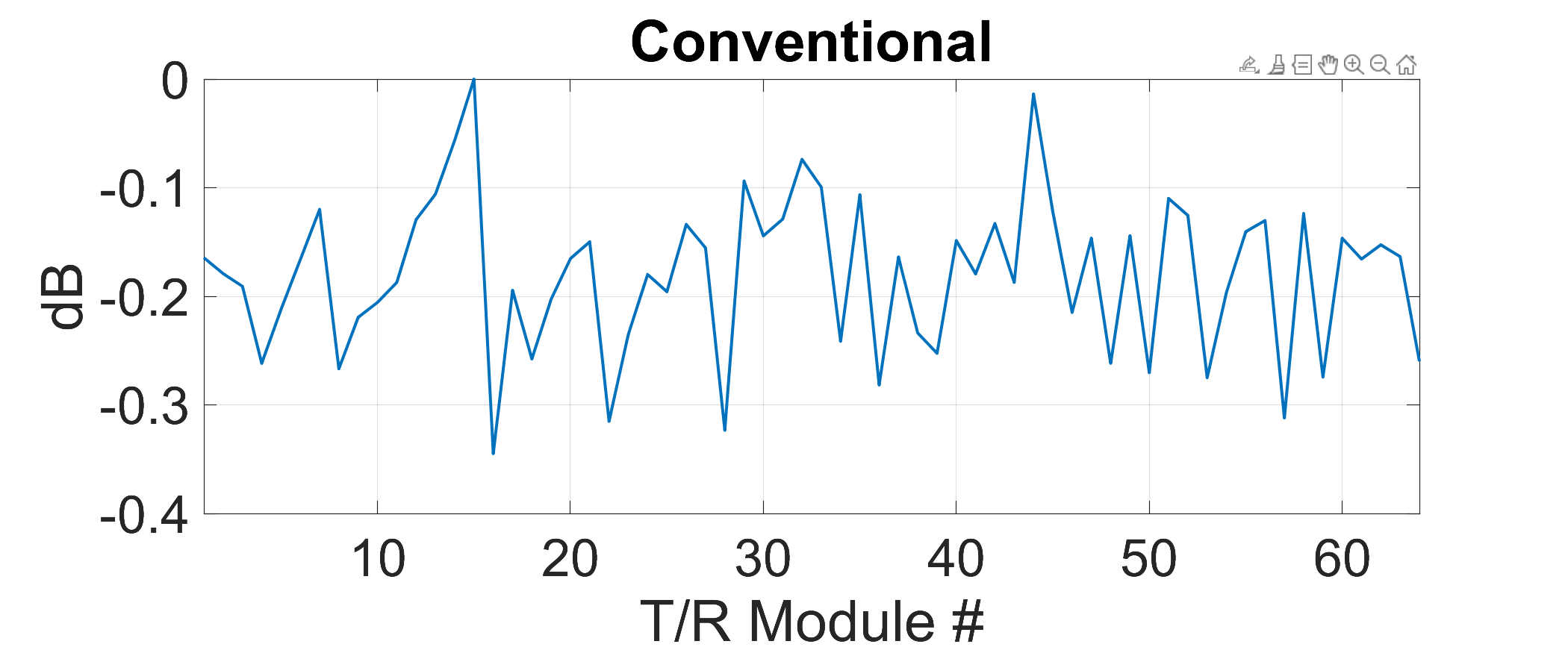}\label{trm_error_conventional_amplitude}}
	\hspace{1em}
	\subfloat[][The phase errors for each TRM by the conventional calibration method.]{\includegraphics[width=6 cm]{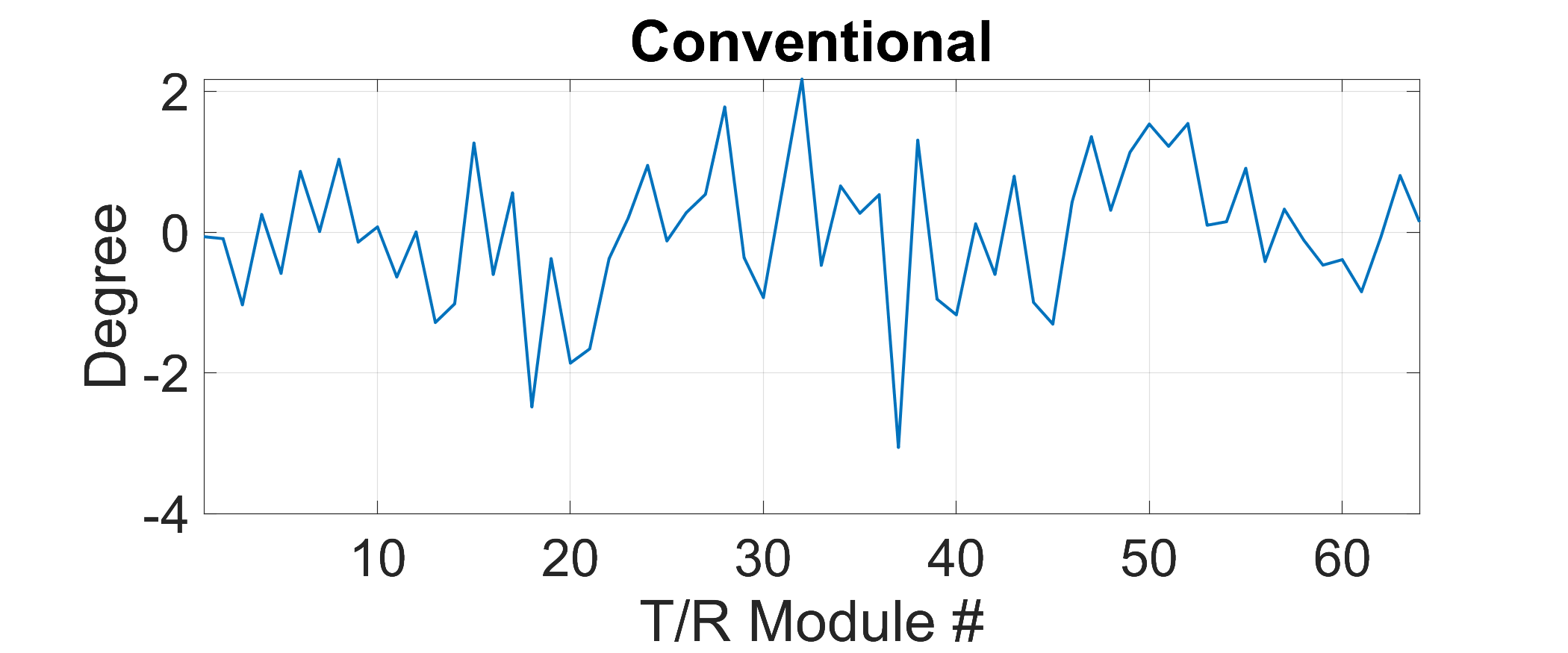}\label{trm_error_conventional_phase}}
	\vspace{0.1em}
	\centering
	\subfloat[][The amplitude errors in terms of frequency for each TRM by the proposed calibration method.]{\includegraphics[width=6 cm]{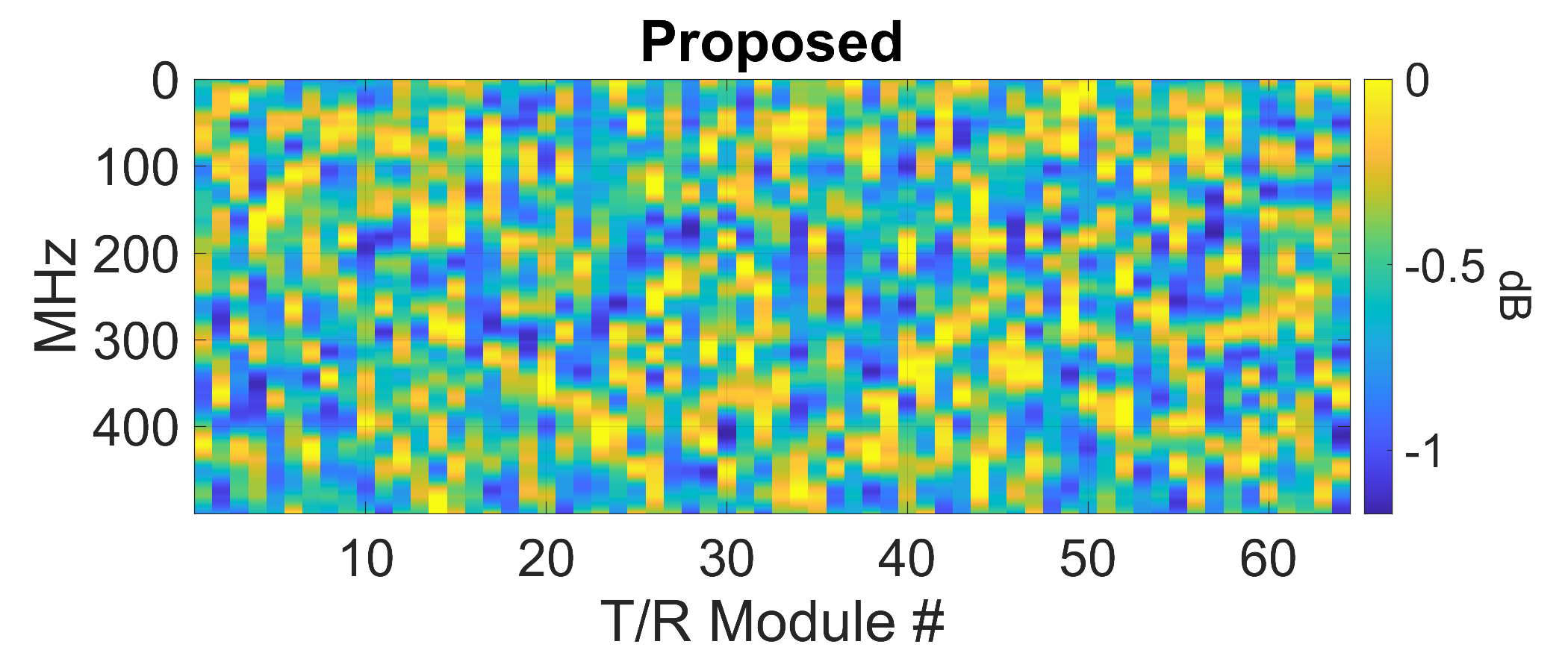}\label{trm_error_proposed_amplitude}}
	\hspace{1em}
	\subfloat[][The phase errors in terms of frequency for each TRM by the proposed calibration method.]{\includegraphics[width=6 cm]{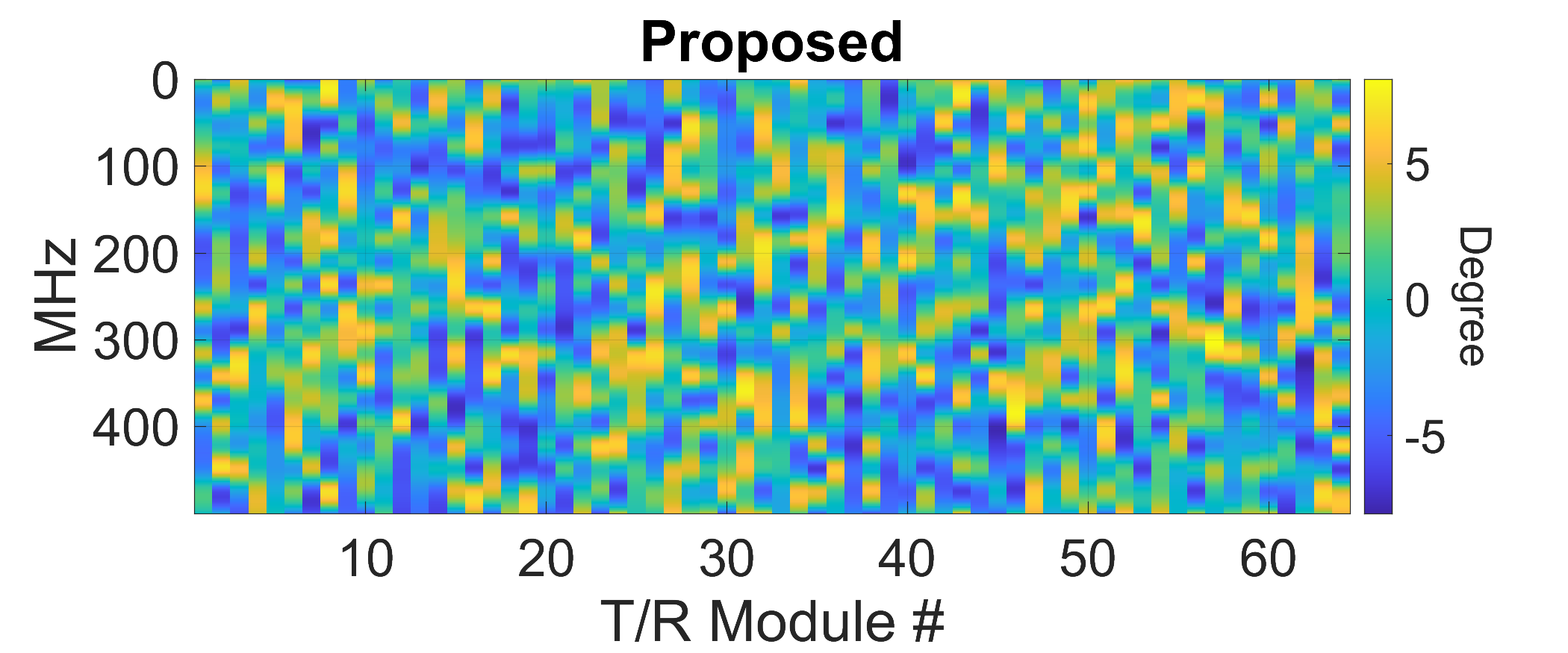}\label{trm_error_proposed_phase}}
	\caption{The calibration result with randomly generated errors for all TRMs. The error values are randomly generated as described in Table. \ref{table_error}.}
	\label{trm_calibration_values}
\end{figure}

In this scenario, all TRMs are assigned with randomly generated errors, and the conventional and proposed calibration method are performed.
The Fig. \ref{trm_calibration_values} shows the error matrix from the conventional and proposed calibration method for all TRMs.
The assigned error values are randomly generated as described in Table. \ref{table_error}.

\begin{figure}[H]
	\subfloat[][]{\includegraphics[width=6 cm]{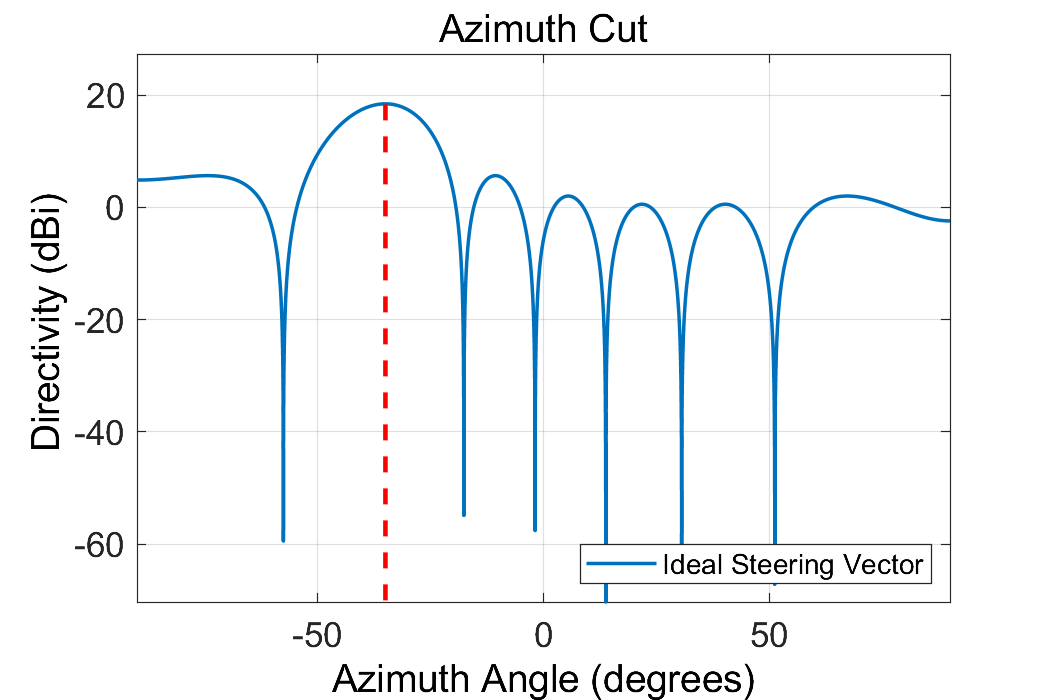}\label{beampattern_ideal}}
	\subfloat[][]{\includegraphics[width=6 cm]{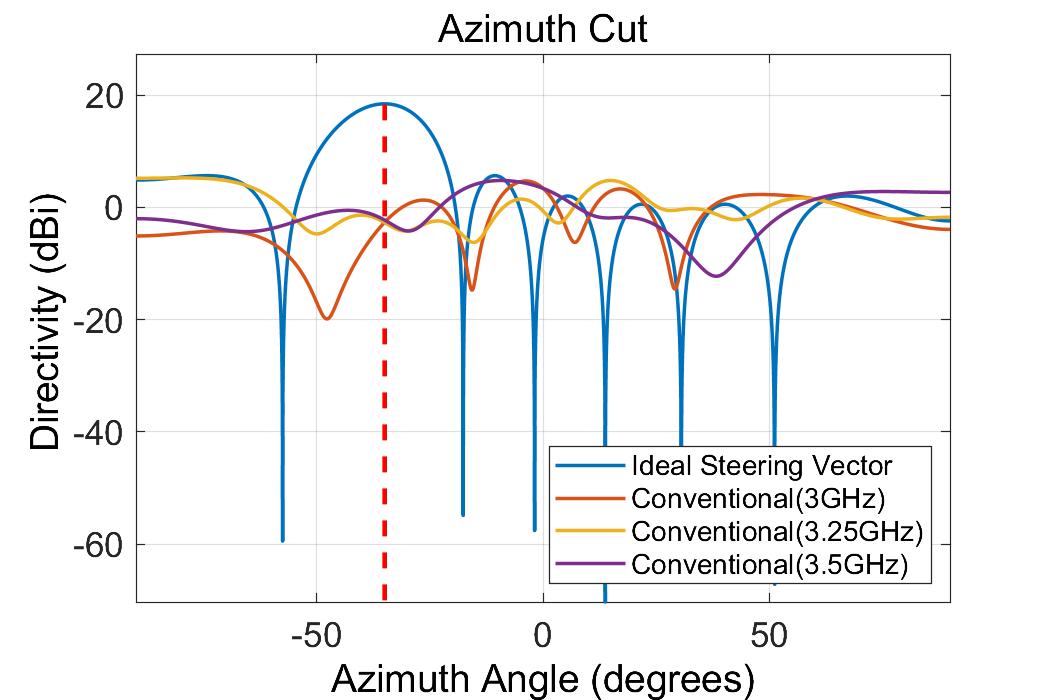}\label{beampattern_conventional}}
	\subfloat[][]{\includegraphics[width=6 cm]{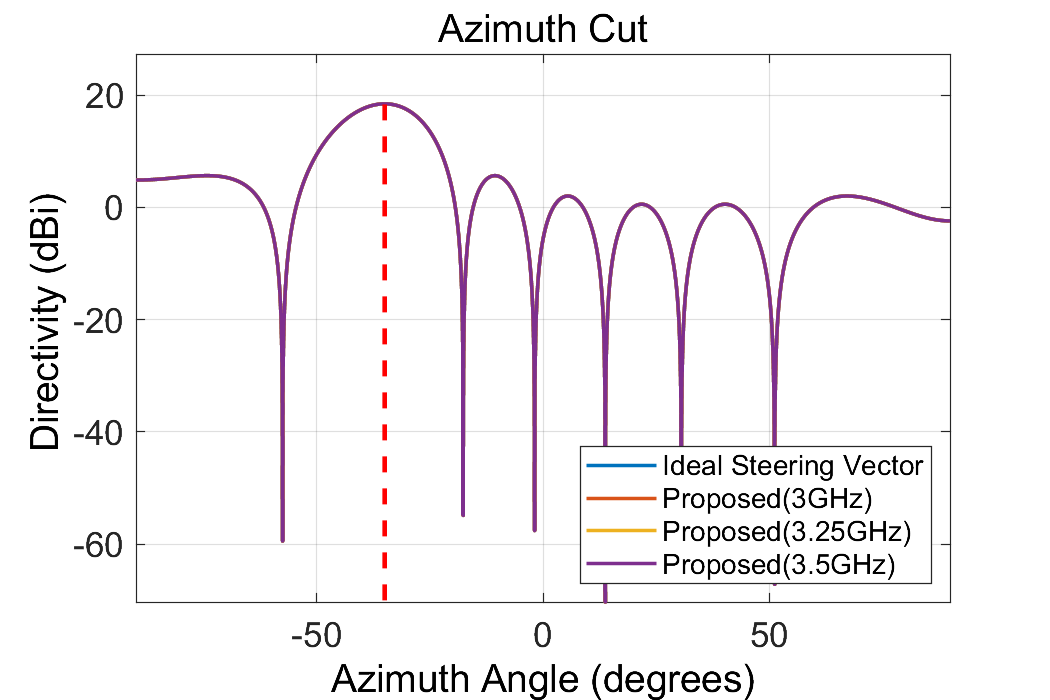}\label{beampattern_proposed}}
	\caption{(a) The desired beamforming result with ideal steering vector. (b) The beamforming result with steering vector calibrated with the conventional method. Due to frequency dependent amplitude and phase error described in Fig. \ref{trm_calibration_values}, the beamforming result is distorted for all LFM frequencies. (c) The beamforming result with steering vector calibrated with the proposed method. Compared to the conventional method, the proposed method can calibrate the frequency dependent amplitude and phase error with time delay error.}
	\label{beampatterns}
\end{figure}

Supposing that the TRMs are performing beamforming in the direction of -35 degree in azimuth plane, the expected beamforming result is presented in the Fig. \ref{beampattern_ideal}.
Based on the measured error value from conventional method and proposed method, the beamforming result is presented in Fig. \ref{beampattern_conventional}, and \ref{beampattern_proposed}.

From the beamforming result, the proposed method can calibrate the frequency dependent amplitude and phase error with time delay error, and the beamforming result is not distorted for all LFM frequencies.
However, the conventional method cannot calibrate the frequency dependent amplitude and phase error, and the beamforming result is distorted for all LFM frequencies.
The main contribution for the beam distortion is the frequency dependent phase error which is from the characteristic of TRM and precise delay error.


\subsection{Discussions}

The proposed method calibrates both precise time delay error and phase error after coarse time delay calibration. 
The impact of the precise time delay error is more significant than the phase error when the sampling rate is low. 
This means that if the sampling rate can be increased, 
the precise time delay error will be reduced because the time delay error can be calibrated at the coarse time delay calibration step.
The higher sampling rate will make coarse time delay calibration more accurate. 
However, the increased sampling rate will also increase the number of samples to be processed, which will increase the computational complexity of the calibration process.

The proposed method generates the error matrix by applying matched filter with sliding window to the error signal.
The computational complexity may be problem when the number of TRMs is large. 
In this case, the sub-band concept can be applied to reduce the computational complexity while maintaining the frequency dependent error monitoring capability, as reported in \cite{deng2011internal}.

\begin{figure}[H]
	\centering
	\includegraphics[width=11 cm]{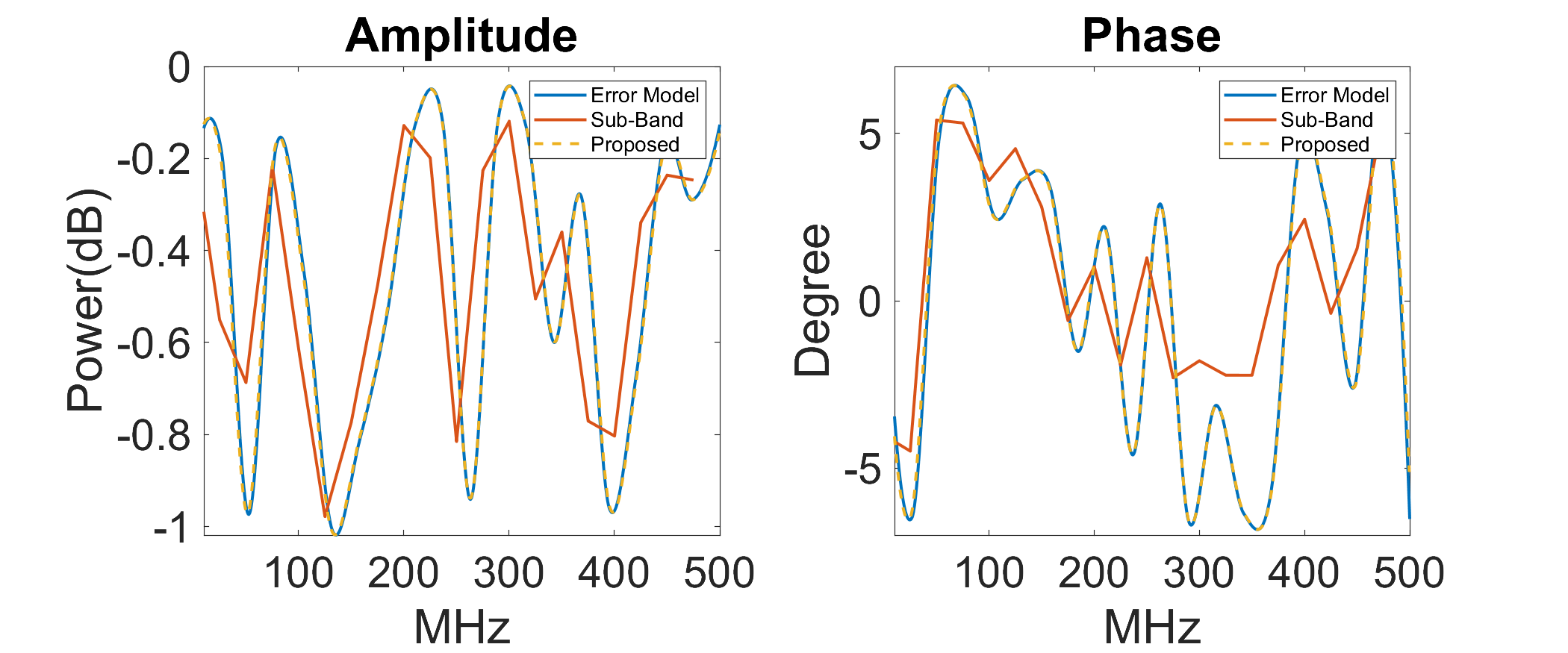}
	\caption{Calibration result comparison between proposed method and sub-band calibration concept. The sub-band matched filtering follows tendency of the frequency dependent error, however, the values are not accurate as the proposed method.  \label{cal_result_comparison_a}}
\end{figure}

The proposed method is compared with the sub-band approach, as shown in Fig. \ref{cal_result_comparison_a}.
The number of sub-band is selected as 20, for 25MHz bandwidth. 
The figure shows that the sub-band matched filtering follows tendency of the frequency dependent error, however, the values are not accurate as the proposed method.
In case that the frequency dependent errors are not changing dramatically, sub-band approach can be applied to reduce the computational complexity.
However, at frequency band where the frequency dependent errors are changing rapidly, 
the sub-band approach cannot be applied because the errors are not accurately measured.
By lowering overlapping ratio of sliding window algorithm, the proposed method can be converted to sub-band approach.
Based on system parameters and signal processing capability, the window size and frequency resolution can be adjusted. 

\section{Conclusion}

The multifunctional and high performance radar system requires the wideband signal to utilize the high range resolution.
The beamforming of wideband signal requires the precise time delay calibration of TTD elements and the frequency dependent amplitude and phase error from the TRM.
As known in literatures, the wideband beamforming requires precise time delay control for each TRM.

This paper presents a new calibration method for the wideband LFM beamforming radar system.
The proposed method can calibrate the frequency dependent amplitude and phase error with precise time delay error without any additional hardware.
By applying the stretched processing to the calibration signal, the proposed method can measure and calibrate the LFM signal with frequency dependent amplitude and phase error with precise time delay error.
The error measurement accuracy and its impact on the beamforming result are analyzed.
The proposed method is compared with true error values and conventional calibration results, and it verified its calibration performance by the simulation results.
The accuracy of the proposed method is determined by following factors: the sampling rate, and the window size and sliding step size of the stretched processing.

The proposed method can be applied to with and without the orthogonal code based calibration method.
This is because the main process of the proposed method that enables frequency dependent monitoring is the sliding window algorithm, which can be applied to any LFM based radar transceivers.
Also, other types of transceivers can adopt the proposed method by generating the LFM signal for calibration purpose.
The proposed method can be extended to the optimization of the parameter selection of the stretched processing for reducing the computational complexity.

\vspace{6pt}


\section*{Acknowledgments}

This work is supported by HANWHA Systems.

\bibliographystyle{IEEEtran}
\bibliography{sliding}

%


\end{document}